# Radiation of a Bunch Crossing a Boundary between a Vacuum and Cold Magnetized Plasma in a Waveguide

**Tatiana Yu. Alekhina and Andrey V. Tyukhtin**

St. Petersburg State University
St. Petersburg State University, 7/9 Universitetskaya nab., St. Petersburg, 199034 Russia

E-mails: t.alehina@spbu.ru, a.tyuhtin@spbu.ru

**Abstract**. We analyze the electromagnetic field of a small bunch that uniformly moves in a circular waveguide and transverses a boundary between an area filled up with cold magnetized electron plasma and a vacuum area. The magnetic field is supposed to be strong but finite so the perturbation technique can be applied. Two cases are studied in detail: the bunch is flying out of the plasma into a vacuum, and, inversely, the bunch is flying into the plasma out of the vacuum area of waveguide. The investigation of the waveguide mode components is performed analytically with methods of the complex variable function theory. The main peculiarities of the bunch radiation in such situations are revealed.

## 1. Introduction

The radiation processes in plasma play an important role in accelerator physics, radioscience, cosmic ray physics, and other areas. The study of the radiation processes from the moving charged particles in such a medium was initiated in the 1940s [1,2]. Various researches (both theoretical and experimental) [3-10] indicated that Vavilov-Cherenkov (or Cherenkov) radiation (CR) generated in plasma can be very perspective. It allows a significant increase in the gain efficiency and in the total power of oscillations, and, moreover, demonstrates the ability by changing the plasma density to tune the frequencies generated Microwave oscillations.

These advantages was also used in the plasma filled waveguides which were of great interest since the 1960s [11-14] when it was shown that such beam-plasma devises could be useful for the purpose of development of high power electromagnetic radiation sources (both nonrelativistic and relativistic) [15-18] and particle acceleration [13, 19-26] including hybrid plasma - dielectric wakefield accelerators (PDWA). Note that as contrast to dielectric wakefield accelerators (DWA) [27, 28] which withstand very high-gradient fields (>1 GV/m) PDWA can also provide an important radial focusing force on the witness bunch.

Results have also been presented in reviews and monographs [4, 7-9, 14, 19], but attention has mainly paid to the analyses of the particular energetic characteristics of radiation, without considering the structure of the field in the plasma. In addition, the effect of the transverse boundary on the wave fields in such structures were not also analyzed. Note that limitation in the length of slowing structures can significantly affect the excitation of wake waves by bunches. The electromagnetic field structure for the case of infinite cold magnetized gyrotropic plasma was investigated in [29] and for the case of semi-infinite anisotropic medium with dispersion of a plasma type was considered in [30,31].

It should be noted that CR can penetrate through the boundary in some situation, and so-called Cherenkov-transition radiation (CTR) can be generated. The main peculiarities of the CTR effect were described for the different semi-infinite media in waveguides (isotropic and anisotropic dielectrics, medium with the dispersion of a resonant type, left-handed medium, strong magnetized plasma, plasma-like media) [32-37].

In this paper, we consider the situation when the magnetic field in plasma is strong enough but finite so the perturbation technique can be applied with a small parameter $h$ which is equal to the ratio of plasma frequency to gyrotropic frequency ($h \ll 1$). Note that in this case the zeroth-order approximation corresponds to the strong magnetic field when the gyration parameter is neglected and the dielectric permittivity tensor is diagonal [37]. Two cases are studied in detail: the bunch is flying out of the plasma into a vacuum, and, inversely, the bunch is flying into the plasma out of the vacuum area of waveguide. The main goal of the paper is to study features of generated radiations in these two cases. The phenomena under consideration are interesting in connection with development of new methods of generation of gigahertz and terahertz radiation as well as for diagnostics of beam and medium characteristics.

## 2. The problem statement

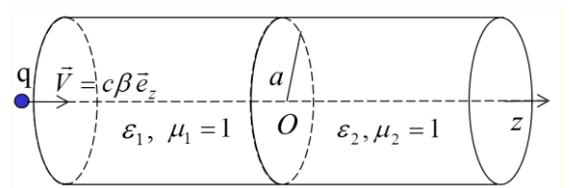

**Figure 1.**(Colour online) Geometry of the problem.

We consider a small bunch $q$ that moves with a constant velocity $\vec{V} = c\beta \vec{e}_z$ ($c$ is the light velocity in vacuum) in a metal circular waveguide of a radius $a$ and crosses the boundary between a vacuum and a homogeneous magnetized cold electron plasma. The magnetic field $H_0$ in plasma is strong enough but finite so the perturbation technique can be applied with a small parameter $h = \omega_p/\omega_H$, $h \ll 1$, where $\omega_p^2 = 4\pi n_p e^2/m$ is the plasma frequency ($n_p$ is the electron density, $e$ and $m$ are the electron charge and the electron mass, respectively), and $\omega_H = |e|H_0/(mc)$ is the gyrotropic frequency. From the point of view of macroscopic electrodynamics, the plasma is characterized by the dielectric permittivity tensor [7]:

$$\hat{\varepsilon} = \begin{pmatrix} \varepsilon_r & -ig & 0 \\ ig & \varepsilon_r & 0 \\ 0 & 0 & \varepsilon_z \end{pmatrix}, \quad (1)$$

with components $\varepsilon_r = 1 + O(h^2)$, $\varepsilon_z(\omega) = 1 - \dfrac{\omega_p^2}{\omega^2 + 2i\omega\nu}$ and $g = h\omega_p/\omega + O(h^2)$, where $\nu$ is the effective collision frequency (we will set $\nu \to +0$ in final results).

The bunch moves along the $z$-axis of the cylindrical frame of reference, $r$, $\varphi$, $z$. It is characterized by some distribution along the $z$-axis and a negligible thickness, i. e., the charge density is $\rho = q\delta(r)\eta(\zeta)/(2\pi r)$, where $\int_{-\infty}^{\infty}\eta(\zeta)d\zeta = 1$, $\zeta = z - ct\beta$. For example, for the bunch with the Gaussian distribution $\eta(\zeta) = \exp(-\zeta^2/(2\sigma^2))/(\sqrt{2\pi}\sigma)$, where $\sigma$ is the half length of the bunch. The middle of the bunch intersects the boundary at the moment $t=0$.

We consider two cases in detail: first, the bunch is flying out of the plasma into a vacuum, and, second, the bunch is flying into the plasma out of the vacuum. The parameters for the areas on the left of boundary ($z < 0$) and on the right ($z > 0$) are labeled with subscript 1 and subscript 2, correspondingly (fig.1).

In the plasma area, the zeroth-order approximation of the perturbation method corresponds to the strong magnetic field when the gyration parameter is neglected and the dielectric permittivity tensor is diagonal (approximation of infinite external magnetic field [37]). For the Fourier components of the field, the problem is formulated in the form:

$$\begin{cases} L_r H_\varphi^{(0)} + k_0^2 H_\varphi^{(0)} + \dfrac{\varepsilon_z}{\varepsilon_r}\dfrac{\partial^2}{\partial z^2}H_\varphi^{(0)} = \dfrac{4\pi}{c}\dfrac{\partial}{\partial r}j_\omega, \\ \left.\dfrac{1}{r}\dfrac{\partial}{\partial r}\left(rH_\varphi^{(0)}\right)\right|_{r=a} = 0 \end{cases} \quad (2)$$

$$E_r^{(0)} = -\dfrac{i}{\varepsilon_r k_0}\dfrac{\partial}{\partial z}H_\varphi^{(0)}, E_z^{(0)} = \dfrac{i}{\varepsilon_z k_0}\left[\dfrac{1}{r}\dfrac{\partial}{\partial r}\left(rH_\varphi^{(0)}\right) - \dfrac{4\pi}{c}j_\omega\right],$$

$$L_r = \dfrac{\partial^2}{\partial r^2} + \dfrac{1}{r}\dfrac{\partial}{\partial r} - \dfrac{1}{r^2},$$

$$j_\omega = \dfrac{q}{(2\pi)^2}\dfrac{\delta(r)}{r}\exp\left(\dfrac{i\omega z}{\beta c}\right)\tilde{\eta}(\omega),$$

where $k_0 = \omega c^{-1}$, $\tilde{\eta}(\omega)$ is the Fourier transform of the longitudinal distribution $\eta(\zeta)$, normalized to the distribution of a point charge (for example, $\tilde{\eta}(\omega) = \exp\left[-\omega^2\sigma^2/(2\beta^2 c^2)\right]$ for the bunch with the Gaussian distribution). So, in the zeroth-order approximation we have TM-polarization only $E_\varphi^{(0)} = H_r^{(0)} = H_z^{(0)} = 0$.

In the first-order approximation, we obtain TE-polarization with the problem statement:

$$\begin{cases} L_r E_\varphi^{(1)} + k_0^2 E_\varphi^{(1)} + \dfrac{\partial^2}{\partial z^2}E_\varphi^{(1)} = -\dfrac{\omega_p}{c\varepsilon_r}\dfrac{\partial}{\partial z}H_\varphi^{(0)}, \\ \left.E_\varphi^{(1)}\right|_{r=a} = 0 \end{cases} \quad (3)$$

$$H_r^{(1)} = \dfrac{i}{k_0}\dfrac{\partial}{\partial z}E_\varphi^{(1)}, H_z^{(1)} = \dfrac{i}{\varepsilon_z k_0}\dfrac{1}{r}\dfrac{\partial}{\partial r}\left(rE_\varphi^{(1)}\right),$$

$$H_\varphi^{(1)} = E_r^{(1)} = E_z^{(1)} = 0.$$

Taking into account the first-order approximation only, for the field components we have:

$$H_\varphi = H_\varphi^{(0)}\left[1 + O(h^2)\right], E_{r,z} = E_{r,z}^{(0)}\left[1 + O(h^2)\right], \quad (4)$$
$$E_\varphi = hE_\varphi^{(1)}\left[1 + O(h^2)\right], H_{r,z} = hH_{r,z}^{(1)}\left[1 + O(h^2)\right]$$

The analytical solution of the problem is given in the form of decomposition in an infinite series of normal modes [4]. For TM-polarization (2), the solution is based on the spectral problem for $H_\varphi^{(0)}$ with eigenfunctions $J_1(\chi_{0n}r/a)$ and eigenvalues

$$\lambda_n = \chi_{0n}/a, \quad J_0(\chi_{0n}) = 0, \quad n = 1, \ldots. \quad (5)$$

For TE-polarization (3), the spectral problem for $E_\varphi^{(1)}$ gives eigenfunctions $J_1(\chi_{1m}r/a)$ and eigenvalues

$$\lambda_m = \chi_{1m}/a, \quad J_1'(\chi_{1m}) = 0, \quad m = 1, \ldots. \quad (6)$$

To obtain the field components for TE-polarization, we use noun integral [38]:

$$\int_0^a rJ_1(\chi_{1m}r/a)J_1(\chi_{0n}r/a)dr = \dfrac{a^2\chi_{1m}}{\chi_{0n}^2 - \chi_{1m}^2}J_0(\chi_{1m})J_1(\chi_{0n}). \quad (7)$$

For the problems with the transverse boundary, the problem statement is supplemented with the boundary conditions at $z=0$:

$$H_{\varphi 1} = H_{\varphi 2}, E_{r1} = E_{r2} \quad \text{(for TM –polarization)} \quad (8)$$

and $E_{\varphi 1} = E_{\varphi 2}, H_{r1} = H_{r2}$ (for TE –polarization).

So, the field components can be written in the form [39]:

$$\vec{H}_{1,2} = \vec{H}_{1,2}^{q} + \vec{H}_{1,2}^{b}, \vec{E}_{1,2} = \vec{E}_{1,2}^{q} + \vec{E}_{1,2}^{b}, \quad (9)$$

where subscript 1 and subscript 2 relate to the areas $z < 0$ and $z > 0$, respectively. The first summands in (9) describe so-called 'forced' field, which is the field of a bunch in a regular waveguide. The forced field can contain CR. The second summands in (9) are the 'free' field components [39] that connected with the influence of the boundary. The free field can contain CTR and transition radiation (TR). We obtain the field components for two different cases, when the bunch is flying out of the plasma into a vacuum, and, inversely, out of the vacuum into the plasma. Further, the field components are analyzed analytically with methods of the complex variable function theory. The main peculiarities of the bunch radiation are revealed in both situations.

## 3. The case of flying out of plasma into a vacuum

### 3.1. General solution

First, we consider the case when the bunch crosses the boundary between the plasma described with by the dielectric permittivity tensor (1) on the left ($z<0$) and a vacuum on the right ($z>0$). We use the perturbation method and found the field in the plasma in the form (4) with Eq. (2) for TM-polarization and Eq. (3) for TE-polarization simultaneously with the boundary conditions (8). Based on spectral problems for TM- and TE-polarizations (5)-(7) we have general solution of the problem in the form (9). Here we only give expression for the $\varphi-$components of the magnetic and electric fields:

$$H_{\varphi 1,2}^q = \frac{-2q}{\pi c a^3}\sum_{n=1}^{\infty}\alpha_n J_1\left(\frac{\chi_{0n}r}{a}\right)\int_{-\infty}^{+\infty}d\omega\,\tilde{\eta}(\omega)h_{n1,2}(\omega)\exp\left[\frac{i\varsigma}{c\beta}\right], \quad (10)$$

$$H_{\varphi 1,2}^b = \frac{-2q\beta}{\pi c a^3}\sum_{n=1}^{\infty}\alpha_n J_1\left(\frac{\chi_{0n}r}{a}\right)\int_{-\infty}^{+\infty}B_{n1,2}\tilde{\eta}(\omega)\exp\left[i(k_{z1,2}|z|-\omega t)\right]d\omega, \quad (11)$$

$$E_{\varphi 1}^q = -\frac{4iqh\omega_p}{\pi\beta c^3 a^3}\sum_{n=1}^{\infty}\sum_{m=1}^{\infty}\alpha_{mn}J_1\left(\frac{\chi_{1m}r}{a}\right)\int_{-\infty}^{+\infty}\frac{\omega d\omega h_{n1}(\omega)\tilde{\eta}(\omega)\exp[i\varsigma/c\beta]}{k_{z0}^{(m)2}-\omega^2(\beta^2c^2)^{-1}}, \quad (12)$$

$$E_{\varphi 2}^q = 0, \quad E_{\varphi 1}^b = E_{\varphi 1}^{b1} + E_{\varphi 1}^{b2},$$

$$\begin{Bmatrix} E_{\varphi 1}^{b1} \\ E_{\varphi 2}^b \end{Bmatrix} = \frac{2iqh\omega_p}{\pi c^2 a^3}\sum_{n=1}^{\infty}\sum_{m=1}^{\infty}\alpha_{mn}J_1\left(\frac{\chi_{1m}r}{a}\right)\int_{-\infty}^{+\infty}d\omega\begin{Bmatrix} e_{mn}^{(1)} \\ e_{mn}^{(2)} \end{Bmatrix}\tilde{\eta}(\omega)\exp\left[i(k_{z0}^{(m)}|z|-\omega t)\right], \quad (13)$$

$$E_{\varphi 1}^{b2} = \frac{4iqh\omega_p}{\pi c^2 a^3}\sum_{n=1}^{\infty}\sum_{m=1}^{\infty}\alpha_{mn}J_1\left(\frac{\chi_{1m}r}{a}\right)\int_{-\infty}^{+\infty}\frac{d\omega\,\tilde{\eta}(\omega)k_{z1}^{(n)}B_{n1}\exp\left[i(k_{z1}^n|z|-\omega t)\right]}{k_{z0}^{(m)2}-\omega^2(\beta^2c^2)^{-1}}, \quad (14)$$

where $\varsigma = z-ct\beta$, $\alpha_n = \chi_{0n}J_1^{-2}(\chi_{0n})$,

$$\alpha_{mn} = \frac{\chi_{0n}\chi_{1m}}{(\chi_{0n}^2-\chi_{1m}^2)J_1(\chi_{0n})J_0(\chi_{1m})},$$

$$h_{n1}(\omega) = \frac{\varepsilon_r}{\varepsilon_z\left[k_{z1}^{(n)2}-\omega^2(\beta^2c^2)^{-1}\right]}, \quad h_{n2}(\omega) = \frac{1}{k_{z2}^{(n)2}-\omega^2(\beta^2c^2)^{-1}},$$

$$k_{z1}^{(n)} = \frac{\omega}{c}\sqrt{\frac{\omega^2-\omega_p^2-\omega_n^2}{\omega^2-\omega_p^2}}, \quad k_{z2}^{(n)} = \frac{1}{c}\sqrt{\omega^2-\omega_n^2}, \quad (15)$$

$$k_{z0}^{(m)} = \frac{1}{c}\sqrt{\omega^2-\Omega_m^2}, \quad (16)$$

$$\omega_n = \chi_{0n}c/a, \quad \Omega_m = \chi_{1m}c/a \quad (17)$$

$$B_{n1,2} = \frac{\omega(\beta c)^{-1}\mp k_{z2,1}^{(n)}}{\varepsilon_z(k_{z1}^{(n)}+k_{z2}^{(n)})}[h_{n1}(\omega)-h_{n2}(\omega)],$$

$$e_{mn}^{(1),(2)} = \mp\frac{k_{z1}^{(n)}B_{n1}}{k_{z0}^{(m)}(k_{z0}^{(m)}\mp k_{z1}^{(n)})} - \frac{\omega(\beta c)^{-1}(\omega(\beta c)^{-1}\mp k_{z0}^{(m)})}{k_{z0}^{(m)}(k_{z0}^{(m)2}-\omega^2(\beta^2c^2)^{-1})}h_{n1}(\omega).$$

Here, it is assumed that $\mathrm{Im}\,k_{z1,2}^{(n)}>0$ (15) and $\mathrm{Im}\,k_{z0}^{(m)}>0$ (16). This condition means that the waves outgoing from the boundary must decrease exponentially with an increase in the distance $|z|$ if we take into account dissipation in the medium. We analyze expressions (10)-(14) using analytical methods based on the complex variable function theory [40] (as it was made in our previous papers [32-37,41] for different problems with sectional homogeneous waveguides).

### 3.2. CR in the plasma area of the waveguide

First, we study the modes of the 'forced' magnetic and electric fields (10) and (12) that is the field of a bunch in a regular waveguide loaded with the plasma under consideration. Note that magnetic field coincides with the one in strongly magnetized cold plasma [37]. Note that CR is not generated in isotropic cold plasma [41] as opposed to the case considered here. The analytical investigation is based on the complex variable function theory [40]. Analysis of the integrands (10), (12) show that there is only one pair of the poles placed at the real

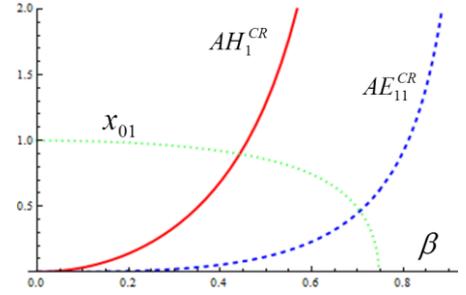

**Figure 2.(Colour online)** Behavior of the CR amplitude of the first modes of TM- (solid red line) and TE-polarization (dashed blue line) and the CR frequency (dotted green line) depending on velocity; $h = 0.2$, $x_p = \omega_p a/c = 2.7$, $\beta_{01} \approx 0.75$.

axis which gives Cherenkov radiation at some condition:

$$\pm\omega_{0n} = \pm\omega_p x_{0n}, \quad x_{0n} = \left[1-\frac{\chi_{0n}^2\beta^2}{x_p^2(1-\beta^2)}\right]^{1/2}, \quad (18)$$

where

$$x_p = \omega_p a/c. \quad (19)$$

The condition for the CR generation is the following:

$$\beta < \beta_{0n} = \left(1+\chi_{0n}^2 x_p^{-2}\right)^{-1/2}. \quad (20)$$

Thus, unlike the case of an isotropic non-dispersive medium [32,34], CR is excited if the charge velocity is less than certain limit value $\beta_{0n}$. Note, that this threshold value decreases with an increase in a mode number $n$.

The frequency of the radiated wave (18) is always less than the plasma frequency $\omega_{0n}<\omega_p$ ($x_{0n}<1$). It should be noted that for this frequency range the group velocities of the radiated waves in such medium is directed towards the bunch motion (positive direction of the axis $z$).

The forced field in the plasma consists of two parts: a quasi-Coulomb field and CR which has the components:

$$H_{\varphi 1}^{CR} = \frac{4q}{a^2}\sum_{n=1}^{\infty}J_1\left(\frac{\chi_{0n}r}{a}\right)\tilde{\eta}(\omega_{0n})AH_n^{CR}\sin\left[\omega_{0n}\left(t-\frac{z}{c\beta}\right)\right]\theta(\beta ct-z),$$

$$E_{\varphi 1}^{CR} = \frac{4q}{a^2} \sum_{n=1}^{\infty} \sum_{m=1}^{\infty} J_1\left(\frac{\chi_{0n} r}{a}\right) \tilde{\eta}(\omega_{0n}) AE_{mn}^{CR} Cos\left[\omega_{0n}\left(t - \frac{z}{c\beta}\right)\right] \theta(\beta ct - z),$$

$$AH_n^{CR} = \frac{\beta^2 \alpha_n}{x_{0n} x_p (1 - \beta^2)}, \quad (21)$$

$$AE_{mn}^{CR} = \frac{2h\beta^3 \alpha_{mn}}{x_{0n}^2 x_p (1 - \beta^2)(\beta^2 \Psi_{mn}^2 - 1)}, \quad (22)$$

where $\Psi_{mn}^2 = 1 - \chi_{1m}^2 (x_p^2 x_{0n}^2)^{-1}$, $\theta(x)$ is the Heaviside step function.

The behavior of the CR amplitudes of both polarizations and the CR frequency for the first mode of the wakefield for different values of $\beta$ is presented in Fig. 2. At low velocities of $\beta$ the CR frequency is about the plasma frequency and the CR amplitudes are relatively small. The CR amplitude increase and the CR frequency tends to zero if $\beta$ tends to the threshold value $\beta_{0n}$. Note that the frequencies can be easily tuned by changing the plasma density $n_{pe}$.

### 3.3. Radiation of the bunch in the plasma area of the waveguide

The investigation for the free field components (11), (13), (14) show a rather complex picture of the integrand singularities on the complex plane of $\omega$. For TM-polarization, the analysis is the same as in [37]. For TE-polarization, in addition to the poles $\pm \omega_{0n}$ (18) (the singularities of the forced field) there are as well:
- the poles

$$\pm \Omega_{0m} = \pm \omega_p y_{mn}, \quad y_{mn} = (1 - \chi_{0n}^2 \chi_{1m}^{-2})^{-1/2}; \quad (23)$$

- the poles on the imaginary axis $\pm \omega_{0n}^{(1)} = \pm i\beta \omega_n (1 - \beta^2)^{-1/2}$ and $\pm \Omega_{0m}^{(1)} = \pm i\beta \Omega_m (1 - \beta^2)^{-1/2}$;
- the branch points of the radicals $k_{z1,2}^{(n)}$ (15) and $k_{z0}^{(m)}$ (16) $\pm \omega_p - i0$, $\pm \tilde{\omega}_n^{(1)} - i0$, $\tilde{\omega}_n^{(1)} = \sqrt{\omega_p^2 + \omega_n^2}$, $\pm \omega_n - i0$ and $\pm \Omega_m - i0$. Disposition of the singularities of the integrands, branch cuts and integration path in a complex plane of $\omega$ is shown in Fig. 3.

It should be noted that the poles $\pm \Omega_{0m}$ are placed at the real axes if the following condition is satisfy:

$$\chi_{1m} - \chi_{0n} > 0. \quad (24)$$

Then these poles give the discrete spectrum of transition radiation (DSTR) so for every nth mode CR there are $m = n, n+1, ...$ modes of the DSTR. As the frequencies of CR are always less than the plasma frequency (the low-frequency branch), the frequencies $\Omega_{0m}$ are always more than the plasma frequency $\Omega_{0m} > \omega_p$ (the high-frequency branch). Moreover, the frequencies $\Omega_{0m}$ do not depend on the velocity of the bunch motion $\beta$. Fig. 3 shows all of the singularities on the complex plane of $\omega$ and the comparative disposition of the initial integration path. The branch cuts defined by the equations

$$\text{Re } k_{z1,2}^{(n)} = 0, \quad \text{Re } k_{z0}^{(m)} = 0.$$

Note that the poles $\pm \omega_{0n}$, $\pm \Omega_{0m}$ and the branch points are slightly shifted downward from the real axis if small losses are taken into account. Therefore, the integration path goes along the real axis above these poles.

The most important results of the investigation for the free field concern the contributions of the poles $\pm \omega_{0n}$ described by Eq. (18) (gives the CTR) and the poles $\pm \Omega_{0m}$ described by Eq. (23) (gives the DSTR). To obtain asymptotic expressions for the free field mode, we can use the steepest descend technique [40]. Note that the procedure is similar to the procedure that was developed for different situations in our previous studies [32-37].

$$H_{\varphi 1}^{CTR} = \frac{4q}{a^2} \sum_{n=1}^{\infty} J_1\left(\frac{\chi_{0n} r}{a}\right) \tilde{\eta}(\omega_{0n}) AH_{n1}^{CTR} Sin\left[\omega_{0n}\left(t - \frac{z}{c\beta}\right)\right] \theta(v_{g1}t - z),$$

$$E_{\varphi 1}^{CTR} = \frac{4q}{a^2} \sum_{n=1}^{\infty} \sum_{m=1}^{\infty} J_1\left(\frac{\chi_{0n} r}{a}\right) \tilde{\eta}(\omega_{0n}) E_{mn1}^{CTR},$$

$$E_{mn1}^{CTR} = AE_{mn1}^{CTR1} Cos\left[\omega_{0n}\left(t - \frac{|z|}{c\beta}\right)\right] \theta(v_{g1}t - |z|) +$$

$$+ AE_{mn1}^{CTR2} Cos\left[\omega_{0n}\left(t - \frac{|z|}{v_f^{mn}}\right)\right] \theta(v_g^{mn} t - |z|),$$

$$E_{\varphi 1}^{DSTR} = \frac{4q}{a^2} \sum_{n=1}^{\infty} \sum_{m=1}^{\infty} J_1\left(\frac{\chi_{0n} r}{a}\right) \tilde{\eta}(\omega_{0n}) AE_{mn1}^{DSTR} Cos\left[\Omega_{0m}\left(t - \frac{|z|}{v_{f1}^m}\right)\right] \theta(v_{g1}^m t - z),$$

$$AH_{n1}^{CTR} = \frac{\beta^2 \alpha_n (1 - \beta \Psi_n)}{x_{0n} x_p (1 - \beta^2)(1 + \beta \Psi_n)}, \quad (25)$$

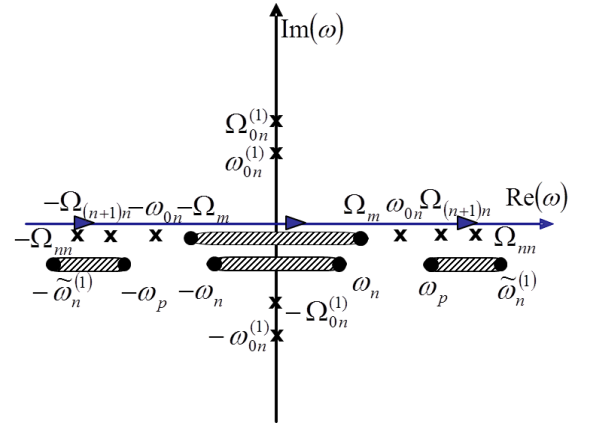

**Figure 3. (Colour online)** Disposition of the singularities of the integrands, branch cuts and integration path in a complex plane of $\omega$ for the nth mode of the free field components in the plasma when condition (20) and (24) are satisfied. The poles and the branch points are shown with crosses and circles, correspondingly.

$$AE_{mn1}^{CTR1} = \frac{2h\beta^3 \alpha_{mn}(1-\beta\Psi_n)}{x_{0n}^2 x_p (1-\beta^2)(\beta^2\Psi_{mn}^2 - 1)(1+\beta\Psi_n)},$$

$$AE_{mn1}^{CTR2} = \frac{2h\beta^3 \alpha_{mn}}{x_{0n}^2 x_p (1-\beta^2)(\beta^2\Psi_{mn}^2 - 1)(1+\beta\Psi_n)}, \quad (26)$$

$$AE_{mn1}^{DSTR} = \frac{2h\beta\beta_p \alpha_{mn}\Psi_m (1-\beta\Psi_{mn}^{(2)})}{(\chi_{0n}^2 - \chi_{1m}^2)(\Psi_m + \Psi_{mn}^{(2)})} \times$$
$$\times \left[\frac{1}{(\beta^2\Psi_m^2 - 1)} - \frac{\chi_{0n}^2}{\chi_{1m}^2(\beta^2\Psi_{mn}^{(2)2} - 1)}\right], \quad (27)$$

where $\Psi_n = \sqrt{1-\chi_{0n}^2(x_p x_{0n})^{-2}}$, $\Psi_m = \sqrt{1-\chi_{1m}^2(x_p y_{mn})^{-2}}$,

$\Psi_{mn} = \sqrt{1-\chi_{1m}^2(x_p x_{0n})^{-2}}$, $\Psi_{mn}^{(2)} = \sqrt{1-\chi_{0n}^2(x_p y_{mn})^{-2}}$,

$v_{g1} = c\beta^{-1}$, $v_f^{mn} = c\Psi_{mn}^{-1}$, $v_g^{mn} = c\Psi_{mn}$, $v_{f1}^m = c\Psi_m^{-1}$, $v_{g1}^m = c\Psi_m$. (28)

In the plasma area of the waveguide, radiation of the bunch has two branches in the frequency band. The low-frequency branch, CR, reflects from the boundary and the CTR is generated. The front of the TM-mode of the CTR propagates with the group velocity $v_{g1}$, the TE-mode of the CTR with numbers $n$ and $m$ consists of two waves that propagate with different velocities $v_{g1}$ and $v_g^{mn}$. CR can also penetrates into a vacuum, as we will see further.

The high-frequency brunch, the DSTR, propagates with the group velocity $v_{g1}^m$ in the direction from the boundary. The frequencies of the radiated waves do not depend on the velocity of the bunch motion $\beta$. However, the amplitudes depend and we can choose an optimal $\beta$ when this amplitude has a maximum value (Fig. 5). Note that the DSTR amplitude is proportional to the magnitude $\approx(\chi_{0n}^2 - \chi_{1m}^2)^{-2}$ and the CTR amplitude is proportional to the magnitude $\approx(\chi_{0n}^2 - \chi_{1m}^2)^{-1}$, so the main contributions in the RSTR give the modes with number $m = n$.

### 3.4. Radiation of the bunch in the vacuum area of the waveguide

Analogous investigation can be made for the free field components (10), (11) and (13) in the vacuum area of the waveguide. Analysis of the integrand singularities on the complex plane of $\omega$ shows that there is only one pair of poles $\pm\omega_{0n}$ (18) which gives the CTR (or the transmitted wave of CR). The DSTR is absent in the vacuum area of the waveguide.

For TM-polarization, if $\omega_n < \omega_{0n}$, then the contributions of the poles $\pm\omega_{0n}$ is a propagating wave, that is, this mode is a part of the CTR. If $\omega_n > \omega_{0n}$, then this mode exponentially decreases with the distance from the boundary and it does not transport the electromagnetic energy.

One can show that the number of propagating modes (which compose the CTR) is always finite. The condition for exciting the propagating mode with the number $n$ is the following: if $\chi_{0n} < x_p$, then the CTR mode with the number $n$ exists in some domain in the vacuum area at the bunch velocity

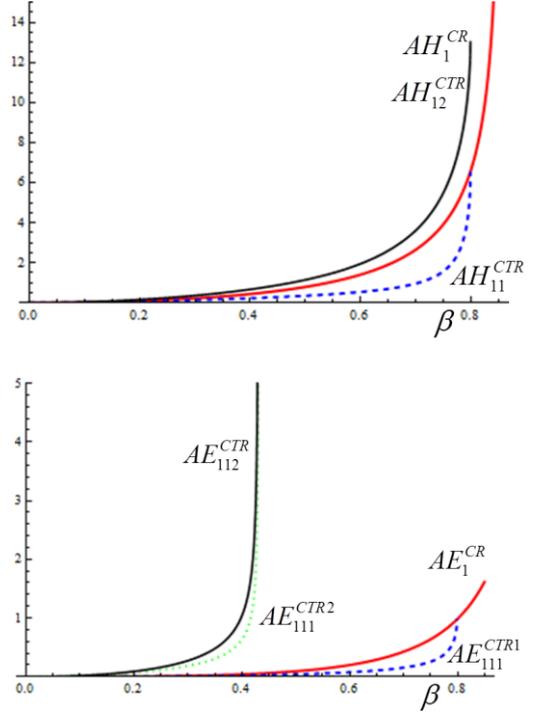

**Figure 4. (Colour online)** Behavior of the amplitudes of the first modes of TM- (top) and TE-polarizations (bottom): CR (solid red line), the CTR in plasma and the CTR in a vacuum (solid black line) depending on velocity; $h = 0.2$, $x_p = \omega_p a/c = 4$, $\beta_{01} = 0.856$, $\beta_{11} = 0.8$ and $\beta_{211} = 0.43$.

$$\beta < \beta_{1n} = (1 - \chi_{0n}^2 x_p^{-2})^{1/2}. \quad (29)$$

For TE-polarization, if $\Omega_m < \omega_{0n}$, then the contributions of the poles $\pm\omega_{0n}$ gives the CTR as well. The condition for exciting the propagating mode with the numbers $n$ and $m$ is the following: if $\chi_{1m} < x_{0n}$, then the CTR mode with the number $n$ exists in some domain in the vacuum area at the bunch velocity

$$\beta < \beta_{2mn} = (1 - \chi_{0n}^2 (x_p^2 - \chi_{1m}^2)^{-1})^{-1/2}. \quad (30)$$

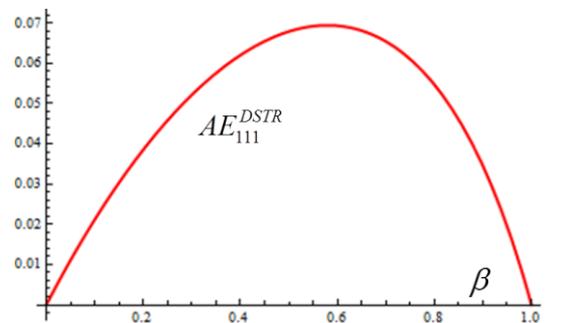

**Figure 5. (Colour online)** The first modes of the DSTR amplitude in the plasma depending on bunch velocity $\beta$; $n = 1$, $m = 1$ and $x_p = 4$.

So, the single TM-mode radiation can be realized for the parameters of the problem that satisfy to the condition:

$$2.405 = \chi_{01} < x_p < \chi_{11} = 3.83, \quad (31)$$

where $x_p$ is nondimensional radius of the waveguide (19). If

$$3.83 = \chi_{11} < x_p < \chi_{02} = 5.52, \quad (32)$$

then the first TM- and TE-modes of the CTR exist in some domain in the vacuum area. For the plasma with $n_{pe} = 10^{14} \text{cm}^{-3}$ (it corresponds to the plasma frequency $\omega_p = 2\pi \cdot 9 \text{GHz}$) the condition (32) gives the limitation of the waveguide radius (in centimeters):

$$2.1 < a < 3. \quad (32a)$$

The threshold values $\beta_{1n}$ and $\beta_{2mn}$ are explained by the total internal reflection of CR off the boundary. Note that the threshold values $\beta_{1n}$ and $\beta_{2mn}$ are always less the CR generation threshold $\beta_{0n}$ (20):

$$\beta_{2mn} < \beta_{1n} < \beta_{0n} \quad (33)$$

The behaviour of the threshold values $\beta_{0n}$, $\beta_{1n}$ and $\beta_{2mn}$ for the first modes of the field is shown in Fig. 5.

Expressions for the CTR in the vacuum area (i.e. transmitted CR) and in the medium (i.e. reflected CR) are as follows:

$$H_{\varphi 2}^{CTR} = \frac{4q}{a^2} \sum_{n=1}^{\infty} J_1\left(\frac{\chi_{0n} r}{a}\right) \tilde{\eta}(\omega_{0n}) AH_{n2}^{CTR} \sin\left[\omega_{0n}\left(t - \frac{z}{v_{f2}}\right)\right] \theta(v_{g2} t - z),$$

$$E_{\varphi 2}^{CTR} = \frac{4q}{a^2} \sum_{n=1}^{\infty}\sum_{m=1}^{\infty} J_1\left(\frac{\chi_{1m} r}{a}\right) \tilde{\eta}(\omega_{0n}) AE_{mn2}^{CTR} \cos\left[\omega_{0n}\left(t - \frac{z}{v_f^{mn}}\right)\right] \theta(v_g^{mn} t - z),$$

$$AH_{n2}^{CTR} = \frac{2\beta^2 \alpha_n}{x_{0n} x_p (1-\beta^2)(1+\beta\Psi_n)}, \quad (34)$$

$$AE_{mn2}^{CTR} = \frac{2h\beta^2 \alpha_{mn}}{x_p \Psi_{mn}(\beta^2\Psi_{mn}^2-1)}\left[1 - \frac{\beta\Psi_n(\beta\Psi_{mn}-1)}{(1+\beta\Psi_n)}\right], \quad (35)$$

where $\Psi_n = \sqrt{1-\chi_{0n}^2(x_p x_{0n})^{-2}}$, $\Psi_{mn} = \sqrt{1-\chi_{1m}^2(x_p x_{0n})^{-2}}$,

$$v_{f2} = c\Psi_n^{-1}, \; v_{g2} = c\Psi_n, \; v_f^{mn} = c\Psi_{mn}^{-1}, \; v_g^{mn} = c\Psi_{mn}. \quad (36)$$

The behavior of the first TM- and TE-modes of the CTR amplitude in a vacuum and the CR amplitude in the plasma at different bunch velocities $\beta$ is presented in Fig. 6. As one can see, the CTR mode amplitude in the vacuum area exceeds the CR mode amplitude. A maximum CTR amplitude is twice as much the CR amplitude.

In the vacuum area, the TM-mode of CTR exists in the domain

$$z < ct\Psi_n, \quad (37a)$$

and the TE-mode of CTR exists in the domain

$$z < ct\Psi_{mn}, \quad (37b)$$

where $\Psi_n$ and $\Psi_{mn}$ are defined by Exp. (36). The inequalities (37) are obtained from the condition of the intersection of the poles at the transformation of the initial integration path to the steepest descent path (analogous procedure was described in [32,34]).

## 4. The case of flying out of a vacuum into the plasma

Second, we consider the case when the bunch is flying out of the vacuum area ($z < 0$) into the plasma described with by the dielectric permittivity tensor (1) ($z > 0$). We use the perturbation method also and found the field in the plasma in the form (4) with Eq. (2) for TM-polarization and Eq. (3) for TE-polarization simultaneously with the boundary conditions (8). We use spectral problems for TM- and TE-polarizations (5)-(7) and obtain general solution of the problem in the form (9). General expressions for the $\varphi$–components of the magnetic and electric fields are the following:

$$H_{\varphi 1,2}^q = \frac{-2q}{\pi c a^3}\sum_{n=1}^{\infty} \alpha_n J_1\left(\frac{\chi_{0n} r}{a}\right)\int_{-\infty}^{+\infty} d\omega \tilde{\eta}(\omega) H_{n1,2}(\omega)\exp\left[\frac{i\zeta}{c\beta}\right], \quad (38)$$

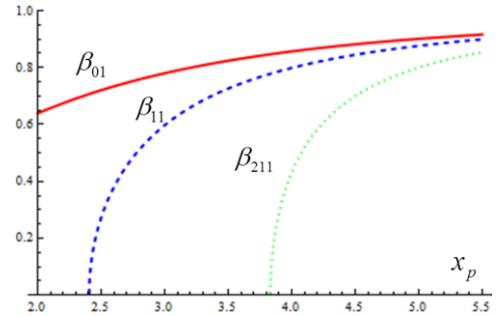

**Figure 6. (Colour online)** The upper threshold values $\beta_{0n}$ (generation of CR) and $\beta_{1n}$ (generation of TM-mode of the CTR) and $\beta_{2nm}$ (generation of TE-mode of the CTR) depending on the parameter $x_p = \omega_p a/c$ for $n = m = 1$.

$$H_{\varphi 1,2}^b = \frac{-2q\beta}{\pi c a^3}\sum_{n=1}^{\infty}\alpha_n J_1\left(\frac{\chi_{0n}r}{a}\right)\int_{-\infty}^{+\infty} A_{n1,2}\tilde{\eta}(\omega)\exp\left[i\left(k_{z1,2}^n|z|-\omega t\right)\right]d\omega, \quad (39)$$

$$E_{\varphi 2}^q = -\frac{4iqh\omega_p}{\pi\beta c^3 a^3}\sum_{n=1}^{\infty}\sum_{m=1}^{\infty}\alpha_{mn}J_1\left(\frac{\chi_{1m}r}{a}\right)\int_{-\infty}^{+\infty}\frac{\omega d\omega H_{n2}(\omega)\tilde{\eta}(\omega)\exp[i\zeta/c\beta]}{k_{z0}^{(m)2}-\omega^2(\beta^2 c^2)^{-1}}, \quad (40)$$

$$E_{\varphi 1}^q = 0, \; E_{\varphi 2}^b = E_{\varphi 2}^{b1} + E_{\varphi 2}^{b2},$$

$$\begin{Bmatrix}E_{\varphi 1}^b \\ E_{\varphi 2}^{b1}\end{Bmatrix} = \frac{2iqh\omega_p}{\pi c^2 a^3}\sum_{n=1}^{\infty}\sum_{m=1}^{\infty}\alpha_{mn}J_1\left(\frac{\chi_{1m}r}{a}\right)\int_{-\infty}^{+\infty}d\omega\begin{Bmatrix}E_{mn}^{(1)} \\ E_{mn}^{(2)}\end{Bmatrix}\tilde{\eta}(\omega)\exp\left[i\left(k_{z0}^{(m)}|z|-\omega t\right)\right], \quad (41)$$

$$E_{\varphi 2}^{b2} = \frac{4iqh\omega_p}{\pi c^2 a^3}\sum_{n=1}^{\infty}\sum_{m=1}^{\infty}\alpha_{mn}J_1\left(\frac{\chi_{1m}r}{a}\right)\int_{-\infty}^{+\infty}\frac{d\omega\tilde{\eta}(\omega)k_{z1}^{(n)}A_{n2}\exp\left[i\left(k_{z1}^n|z|-\omega t\right)\right]}{k_{z0}^{(m)2}-\omega^2(\beta^2 c^2)^{-1}}, \quad (42)$$

where $\varsigma = z - ct\beta$, $\alpha_n = \chi_{0n} J_1^{-2}(\chi_{0n})$,

$$\alpha_{mn} = \frac{\chi_{0n}\chi_{1m}}{(\chi_{0n}^2 - \chi_{1m}^2)J_1(\chi_{0n})J_0(\chi_{1m})},$$

$$H_{n1}(\omega) = \frac{1}{k_{z1}^{(n)2}-\omega^2(\beta^2 c^2)^{-1}}, \; H_{n2}(\omega) = \frac{\varepsilon_r}{\varepsilon_z\left[k_{z2}^{(n)2}-\omega^2(\beta^2 c^2)^{-1}\right]},$$

$$k_{z1}^{(n)} = \frac{1}{c}\sqrt{\omega^2-\omega_n^2}, \; k_{z2}^{(n)} = \frac{\omega}{c}\sqrt{\frac{\omega^2-\omega_p^2-\omega_n^2}{\omega^2-\omega_p^2}}, \quad (43)$$

$$k_{z0}^{(m)} = \frac{1}{c}\sqrt{\omega^2-\Omega_m^2}, \quad (44)$$

$$A_{n1,2} = \frac{\mp\left[k_{z1,2}^{(n)}\pm\omega(\beta c)^{-1}\right]}{\varepsilon_z\left(k_{z1}^{(n)}+k_{z2}^{(n)}\right)}[H_{n1}(\omega)-H_{n2}(\omega)],$$

$$E_{mn}^{(1),(2)} = \mp\frac{k_{z1}^{(n)}A_{n2}}{k_{z0}^{(m)}\left(k_{z0}^{(m)}\mp k_{z1}^{(n)}\right)} - \frac{\omega(\beta c)^{-1}\left(\omega(\beta c)^{-1}\mp k_{z0}^{(m)}\right)}{k_{z0}^{(m)}\left(k_{z0}^{(m)2}-\omega^2(\beta^2 c^2)^{-1}\right)}H_{n2}(\omega).$$

It is assumed that $\text{Im} k_{z1,2}^{(n)} > 0$ (43) and $\text{Im} k_{z0}^{(m)} > 0$ (44) as well as in the first case. To obtain asymptotic expressions for the free field modes, we can also use the steepest descend technique. Here, we only give the results for radiation of the bunch in such situation.

In the vacuum area of the waveguide, both CR and the CTR are absent, and we have the DSTR of TE-polarization at certain condition only.

$$E_{\varphi 1}^{DSTR} = \frac{4q}{a^2}\sum_{n=1}^{\infty}\sum_{m=1}^{\infty} J_1\left(\frac{\chi_{0n}r}{a}\right)\tilde{\eta}(\omega_{0n})AE_{mn1}^{DSTR} Cos\left[\Omega_{0m}\left(t-\frac{|z|}{v_{f1}^m}\right)\right]\theta(v_{g1}^m t - z),$$

$$AE_{mn1}^{DSTR} = \frac{2h\beta\beta_p \alpha_{mn}\Psi_m\left(1+\beta\Psi_{mn}^{(1)}\right)}{(\chi_{0n}^2 - \chi_{1m}^2)(\Psi_m + \Psi_{mn}^{(1)})} \times$$
$$\times\left[\frac{\chi_{0n}^2}{\chi_{1m}^2(\beta^2\Psi_{mn}^{(1)2}-1)} - \frac{1}{(\beta^2\Psi_m^2-1)}\right], \quad (45)$$

where $v_{f1}^m = c\Psi_m^{-1}$, $v_{g1}^m = c\Psi_m$, $\Psi_{mn}^{(1)} = \sqrt{1-\chi_{0n}^2(x_p y_{0m})^{-2}}$,

$$\Psi_m = \sqrt{1-\chi_{1m}^2(x_p y_{0m})^{-2}}. \quad (46)$$

The condition for exciting the propagating TE-mode with the numbers $n$ and $m$ is the following:

$$\sqrt{\chi_{1m}^2 - \chi_{0n}^2} < x_p. \quad (47)$$

So, this condition does not depend on the velocity of the bunch motion but the parameters of the problem only. The single TE-mode of the DSTR ($m = n = 1$) can be realized for the parameters of the problem that satisfy to the condition:

$$2.98 = \sqrt{\chi_{11}^2 - \chi_{01}^2} < x_p, \quad (48)$$

In the plasma area of the waveguide, there are CR, the CTR and the DSTR. Expressions for the $\varphi$ – components of the magnetic and electric fields are the following:

$$H_{\varphi 2}^{CR} = \frac{4q}{a^2}\sum_{n=1}^{\infty} J_1\left(\frac{\chi_{0n}r}{a}\right)\tilde{\eta}(\omega_{0n})AH_n^{CR} Sin\left[\omega_{0n}\left(t-\frac{z}{c\beta}\right)\right]\theta(\beta ct - z),$$

$$E_{\varphi 2}^{CR} = \frac{4q}{a^2}\sum_{n=1}^{\infty}\sum_{m=1}^{\infty} J_1\left(\frac{\chi_{0n}r}{a}\right)\tilde{\eta}(\omega_{0n})AE_{mn}^{CR} Cos\left[\omega_{0n}\left(t-\frac{z}{c\beta}\right)\right]\theta(\beta ct - z),$$

$$AH_n^{CR} = \frac{\beta^2 \alpha_n}{x_{0n}x_p(1-\beta^2)}, \quad (49)$$

$$AE_{mn}^{CR} = \frac{2h\beta^3 \alpha_{mn}}{x_{0n}^2 x_p(1-\beta^2)(\beta^2\Psi_{mn}^2-1)}, \quad (50)$$

$H_{\varphi 2}^{CTR} = -H_{\varphi 2}^{CR}\theta(v_{g2}t-z)$, $E_{\varphi 2}^{CTR} = -E_{\varphi 2}^{CR}\theta(v_{g2}t-z)$,

$$E_{\varphi 2}^{DSTR} = \frac{4q}{a^2}\sum_{n=1}^{\infty}\sum_{m=1}^{\infty} J_1\left(\frac{\chi_{0n}r}{a}\right)\tilde{\eta}(\omega_{0n})AE_{mn2}^{DSTR} Cos\left[\Omega_{0m}\left(t-\frac{|z|}{v_{f2}^m 22}\right)\right]\theta(v_{g2}^m t - z),$$

$$AE_{mn2}^{DSTR} = AE_{mn1}^{DSTR}, \quad (51)$$

where

$v_{g2} = c\beta^{-1}$, $v_{f2}^m = c\Psi_m^{-1}$, $v_{g2}^m = c\Psi_m$, $y_{mn} = (1-\chi_{0n}^2\chi_{1m}^{-2})^{-1/2}$,

$$\Psi_{mn}^{(1)} = \sqrt{1-\chi_{0n}^2(x_p y_{mn})^{-2}}, \quad \Psi_m = \sqrt{1-\chi_{1m}^2(x_p y_{mn})^{-2}}. \quad (53)$$

So, we can see that the CTR in plasma compensates CR in some domain near the boundary $z < v_{g2}t$. The same effect takes place when the bunch is crossing the boundary with some non-dispersive medium as it was shown in our previous works [32,34]. The compensation domain decreases with an increase in $\beta$.

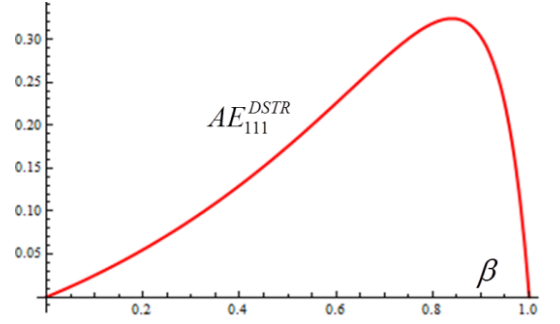

**Figure 7. (Colour online)** The first modes of the DSTR amplitude in a vacuum depending on bunch velocity $\beta$; $n = 1$, $m = 1$ and $x_p = 4$.

The DSTR the numbers $n$ and $m$ are propagating waves if $\chi_{1m} < x_{0n}$, then the DSTR mode exists in some domain near the boundary $|z| < v_{g2}^m t$ where $v_{g2}^m$ is defined by Eq. (53). In a vacuum, the DSTR modes propagate in backward direction of the bunch motion. The frequencies of the radiated waves do not depend on the velocity of the bunch motion $\beta$ and the amplitudes depend, so, we can choose an optimal $\beta$ when this amplitude has a maximum value (Fig. 7). Note that the DSTR amplitude is proportional to the magnitude $\approx (\chi_{0n}^2 - \chi_{1m}^2)^{-2}$ and the CTR amplitude is proportional to the magnitude $\approx (\chi_{0n}^2 - \chi_{1m}^2)^{-1}$, so the main contributions in the RSTR give the modes with number $m = n$.

## 1. Conclusion

The electromagnetic field of a small bunch that moves in a circular waveguide and intersects the boundary between the magnetized cold electron plasma and the vacuum area was considered. The plasma is placed in the magnetic field, which is strong enough, but finite so the perturbation technique can be applied with a small parameter $h$ that is equal to the ratio of the plasma frequency to the gyrotropic frequency ($h \ll 1$). Note that the zeroth-order approximation corresponds to the infinite magnetic field when the gyration parameter is neglected and gives modes of TM-polarization. We restrict the first-order approximation that gives modes of TE-polarization. Two cases are studied in detail: the bunch is flying out of the plasma into a vacuum, and, inversely, the bunch is flying into the plasma out of the vacuum area of waveguide.

In the both cases, radiation of the bunch has two branches in the frequency band. The low-frequency branch, Cherenkov radiation (CR) and the Cherenkov-transition radiation (CTR), are generated in the plasma area at certain condition. The frequencies of CR are always less than the plasma frequency. We have CR of both TM- and TE-polarizations. The high-frequency brunch corresponds to TR and gives the discrete spectrum of transition radiation (DSTR). These waves propagate in the directions from the boundary. Their frequencies are always more than the plasma frequency and do not depend on the velocity of the bunch motion $\beta$. These waves have TE-polarization only.

In the first case when the bunch flies out of the plasma into the vacuum area of the waveguide, several first modes (or only the first mode) of CR can be transmitted through the boundary, and the finite (or single) mode CTR effect takes place in the vacuum area. The other modes of the CR are totally reflected off the boundary. The parameters of the problem can be chosen so that the first TM-mode only or the first modes of both polarizations are generating in a vacuum. The CTR can be the main part of wave field in the vacuum area of the waveguide under some condition. Amplitudes of the CTR modes in vacuum can be greater than the amplitudes of the CR modes in the plasma. The fronts of the TM- and TE-modes of the CTR propagate with different group velocities.

In the second case when the bunch flies out of a vacuum into the plasma, the CTR compensates CR in some domain near the boundary. CR and the CTR propagate in the same direction as the bunch as opposed to the DSTR, which goes in two directions from the boundary (in a vacuum and in the plasma) with the equal group velocity and equal amplitudes. The frequencies of the DSTR do not depend on the velocity of the bunch motion $\beta$. However, the amplitudes depend and we can choose an optimal $\beta$ when this amplitude has a maximum value.

The situations under consideration are effective for generation of radiation from nonrelativistic bunches. The CTR and the DSTR effects can be used for generation of the single-mode monochromatic radiation in the GHz and THz regions in the vacuum parts of the waveguide. The frequencies of generated waves can be tuned by changing the plasma density. It should be also noted that the CTR characteristics can be easily calculated with rather simple formulas without any complicated integrations.

**Acknowledgments**
This research was supported by the Russian Science Foundation (Grant No. 18-72-10137).